\documentclass[10pt, conference]{IEEEtran}
\usepackage{setspace}
\usepackage[utf8]{inputenc}
\usepackage[dvips]{graphicx}
\ifCLASSOPTIONcompsoc
\usepackage[caption=false,font=normalsize,labelfont=sf,textfont=sf]{subfig}
\else 
\usepackage[caption=false,font=footnotesize]{subfig}
\fi
\usepackage{amssymb}
\usepackage[cmex10]{amsmath}
\usepackage{xcolor}
\usepackage{float}
\usepackage{cite}
\usepackage{booktabs}
\usepackage{amssymb}
\usepackage{multirow}
\usepackage{stfloats}
\usepackage{algorithm}
\usepackage{algpseudocode}
\newcommand{\subsubsubsection}[1]{\paragraph{#1}\mbox{}}
\usepackage{microtype}

\usepackage{etoolbox}
\makeatletter
\patchcmd{\@makecaption}
{\\}
{.\ }
{}
{}

\graphicspath{{./figure/}}

\ifCLASSINFOpdf
\else
\fi

\makeatletter
\def\hlinew#1{%
	\noalign{\ifnum0=`}\fi\hrule \@height #1 \futurelet
	\reserved@a\@xhline}
\makeatother%

\allowdisplaybreaks[4]
\usepackage{hyperref}
\usepackage{enumitem}
\setlist[itemize]{leftmargin=*, noitemsep, topsep=0pt}
\setlist[enumerate]{leftmargin=*, noitemsep, topsep=0pt}

\begin{document}

	\title{EdgeCoInfer: Hierarchical Collaborative Inference for On-Device Multimodal Large Models \vspace{-0.2in}} 
	\author{
		Lin Tan\IEEEauthorrefmark{2}\IEEEauthorrefmark{3},
		Songtao Guo\IEEEauthorrefmark{3},
		Mingyan Li\IEEEauthorrefmark{3},
		David K. Y. Yau\IEEEauthorrefmark{2}, 
		\\
		\IEEEauthorblockA{
			\IEEEauthorrefmark{2} Singapore University of Technology and Design\\
			\IEEEauthorrefmark{3} Chongqing University
		}
		\vspace{-0.5in}
	}

	\maketitle
	\thispagestyle{empty}
	
	\begin{abstract}
	To deliver ubiquitous intelligence, modern mobile applications increasingly execute concurrent Multimodal Large Language Models (MLLMs) on edge devices, presenting severe challenges under multi-task concurrency and tight resource constraints. To address this, we propose EdgeCoInfer, a hierarchical collaborative inference framework enabling efficient on-device MLLM inference through coarse-to-fine orchestration. Coarsely, EdgeCoInfer decomposes MLLMs into functional modules for inter-task sharing, avoiding redundant model loading. Finely, it partitions models at the neural network layer level and distributes segments across devices and servers. We jointly optimize layer partitioning, module sharing, and resource allocation under tight constraints. To tackle the non-differentiable combinatorial explosion, we propose a Hybrid Evolutionary Hierarchical Reinforcement Learning (HE-HRL) framework. HE-HRL synchronizes a gradient-free genetic algorithm for discrete partitioning and sharing decisions with a gradient-based soft actor-critic agent for continuous resource refinement. We further embed a constructive cut-step decoder with pre-act pruning and a two-phase curriculum to improve feasibility and accelerate convergence. Experimental results show that EdgeCoInfer breaks the edge memory wall and prevents catastrophic out-of-memory and task failures under high concurrency, reducing memory demand by 53.53\% and system cost by 59.86\% compared to existing methods.

	\end{abstract}

	\begin{IEEEkeywords}
		Edge Artificial Intelligence, Edge Computing, Multimodal Large Language Models, Hierarchical Collaborative Inference
	\end{IEEEkeywords}

	\section{Introduction}
	\subsection{Background and Challenges}
	The surge of large language models (LLMs) and multimodal models marks a major milestone in artificial intelligence (AI)~\cite{10335918,li2025collaborative}. At present, these models are predominantly deployed in centralized cloud infrastructures, a paradigm that introduces several challenges, including excessive end-to-end latency, high bandwidth costs, and heightened data-privacy risks~\cite{10304187,li2026efficient}. A recent shift seeks to place models directly on edge devices (EDs), enabling computation closer to where data are generated~\cite{11183748}. By reducing reliance on remote transmission, this approach can substantially decrease communication overhead and improve responsiveness.
	
	However, the practical deployment of large-scale on-device models faces two fundamental tensions. First, many state-of-the-art AI models are extremely large, with parameter counts that far exceed the memory capacity of any single ED, making local loading and execution infeasible~\cite{10025811}. Second, modern EDs are often required to run multiple AI applications concurrently (e.g., voice assistants, real-time object detection, and AI-enhanced photography). Naively loading a full model for each application not only exhausts scarce static memory resources but also creates severe runtime vulnerabilities due to strictly unshareable, dynamically expanding Key-Value (KV) caches~\cite{zhang2025appagent}. Consequently, orchestrating collaborative inference across multiple edge devices to support heavy-duty multi-task workloads under such strict memory bounds and dynamic KV-cache growth presents a formidable challenge.
	
	\subsection{Motivation and Research Challenges}
	
Existing edge collaborative inference has evolved along two isolated directions. First, Intra-model partitioning distributes a single monolithic model across multiple devices \cite{11571521,11044734,mohammed2020distributed,li2024distributed,borzunov2023distributed,patel2024splitwise,zhang2024edgeshard} to optimize single-task performance. However, applying it to concurrent multi-task scenarios leads to severe resource redundancy and communication overhead. On the other hand, inter-model sharing reuses common functional modules (e.g., encoders) across tasks \cite{han2021legodnn,sun2020adashare,yoon2025s2m3} to conserve memory. Yet, this strategy cannot fit the memory footprints of edge LLMs due to a single task-specific module (e.g., a 7B-parameter LLM head) exceeds the memory capacity of any individual edge device. 
	
To support complex concurrent AI applications, we propose EdgeCoInfer, an efficient hierarchical collaborative inference system. By unifying coarse-grained module sharing with fine-grained layer partitioning, its central brain performs adaptive joint optimization. However, building this system requires overcoming three intrinsically interdependent challenges:

\begin{itemize}

\item Trade-off between Efficiency and Performance: Inter-model sharing minimizes memory footprint but may introduce computation bottlenecks. Conversely, intra-model partitioning accelerates inference but incurs communication overheads \cite{11183748}. Navigating this multi-objective trade-off across heterogeneous edge nodes is non-trivial.
	
\item  Hierarchical Resource Contention: Concurrent multitasking induces multi-granularity conflicts \cite{dahshan2026swarm,jeon2019analysis}. A single device's resources might be simultaneously requested as a standalone host for a complete shared module (coarse-grained) and as a distributed worker for a fraction of a partitioned model (fine-grained), complicating scheduling. 
	
\item  Intractability of the Solution Space: Jointly optimizing sharing, partitioning, and resource allocation creates a combinatorial explosion \cite{9530374}. Coupled with the linear KV-cache growth of MLLMs, this causes severe out-of-memory (OOM) risks and extremely sparse feasible regions under strict memory-latency constraints.

\end{itemize}

	\subsection{Contributions}
	
	To address these multifaceted challenges, we propose EdgeCoInfer, a novel hierarchical collaborative inference framework for on-device MLLMs. The primary contributions of this paper can be summarized as follows.
	\begin{itemize}
		\item We propose a novel hierarchical collaborative inference architecture that enables granularity-adaptive model deployment and inference. Concurrent MLLM deployment and inference are formulated as a joint optimization problem coupling layer-level partitioning, functional-module sharing, and resource allocation under memory, computational, and deadline constraints.
		
		\item We design a problem-structured solution, a Hybrid Evolutionary Hierarchical Reinforcement Learning (HE-HRL) framework, to address the non-differentiable combinatorial explosion problem. Rather than conventional decoupling, a gradient-free genetic algorithm leverages population-based global search over the discrete partitioning and sharing space, avoiding high-variance gradient estimation, while a gradient-based soft actor-critic agent utilizes maximum-entropy exploration for continuous resource refinement.

		
		\item We conduct extensive profile-driven simulations across heterogeneous edge-device capacities and increasing task concurrency. EdgeCoInfer sustains a higher task completion rate under heavy workloads, reduces system cost by 59.86\%, and reduces static and dynamic memory demands by 46.53\% and 70.46\%, respectively.
		

	\end{itemize}


	\section{System Model and Problem Formulation}\label{sec:System Model and Problem Formulation}
	This section presents the system, cost models, and optimization problem of EdgeCoInfer.
	
	\subsection{System Model}
	\subsubsection{System Overview}

\begin{figure*}[t]
		\makebox[\textwidth][c]{\includegraphics[width=0.7\textwidth]{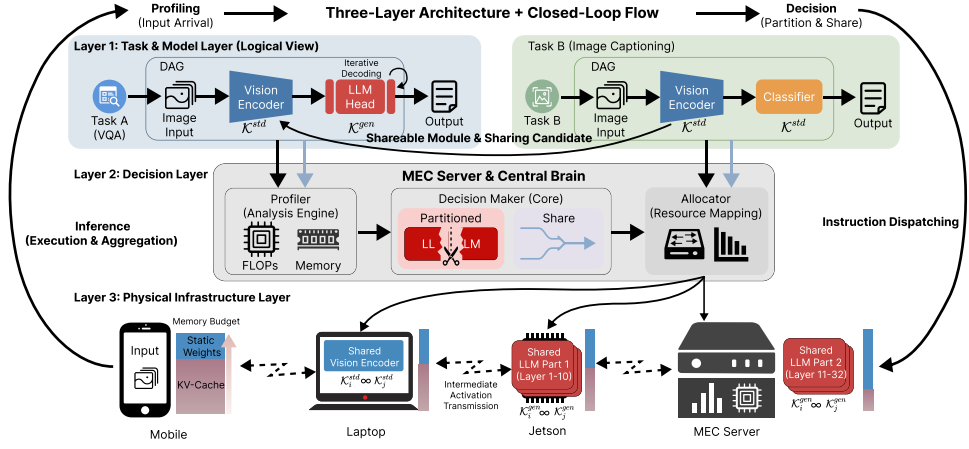}}
		\vspace{-2em}
		\caption{Overview of the EdgeCoInfer framework, which features a three-layer architecture governed by a continuous closed-loop flow. The process commences at the Task and Model Layer, where concurrent multimodal tasks are abstracted into DAGs comprising potentially shareable modules. These requirements are transmitted to the Decision Layer, where a profiler characterizes the workload in terms of FLOPs and memory usage and a decision engine jointly optimizes the partitioning and sharing strategies. Subsequently, the Allocator enforces the resource distribution, and instructions are dispatched to the Physical Infrastructure Layer. Then, devices execute the assigned workloads, and the cycle concludes with inference aggregation as intermediate activations are transmitted to consolidate the final results.}
		\label{fig:1}
	\end{figure*}

    As illustrated in Fig.~\ref{fig:1}, the proposed EdgeCoInfer framework decomposes complex multimodal models into shareable and partitionable functional modules, explicitly categorized into standard modules (e.g., vision encoders) and generative modules (e.g., LLM heads). To enable flexible distributed inference across heterogeneous EDs under strict memory constraints, the system operates on a three-layer architecture governed by a continuous closed-loop flow. 
	
	\subsubsection{System Components and Notation}
	The system comprises one Mobile Edge Computing (MEC) server and $N$ EDs. We denote $\mathcal{N} = \{0,1,2,\ldots,n,\ldots,N\}$ as the set of compute nodes, where $0$ represents the MEC server. 
	
	\subsubsubsection{Device Resource}Each node $n \in \mathcal{N}$ is characterized by the maximum operating frequency $f_n^{max}$ of its on-chip GPU or Neural Processing Unit (NPU), and its memory budget $M_n^{max}$. Considering System-on-Chip architectures where CPUs, GPUs, or NPUs access a unified memory, we do not model host-to-device PCIe transfer overheads. Remaining on-chip memory effects, including bandwidth contention, are implicitly captured within our execution-time and energy models.

	\subsubsubsection{Inference Task}We denote $\mathcal{Q} = \{1, \dots, Q\}$ as the set of concurrent multimodal inference tasks. Let $\mathcal{K}$ denote the universal set of all unique functional module types. Distinct from traditional linear models \cite{mohammed2020distributed}, each task $q \in \mathcal{Q}$ is modeled as a Directed Acyclic Graph (DAG) $G_q = (\mathcal{K}_q, \mathcal{E}_q)$.
	\begin{itemize}
		\item Functional Modules ($\mathcal{K}_q$): The nodes $\mathcal{K}_q \subseteq \mathcal{K}$ denote modules required by task $q$. Identical module types can share static weights across tasks. To capture distinct execution patterns, $\mathcal{K}_q$ is disjointly partitioned into $\mathcal{K}_q^{std}$ and $\mathcal{K}_q^{gen}$. $\mathcal{K}_q^{std}$ includes modules executing a single forward pass (e.g., Convolutional Neural Network (CNN), Vision Transformer (ViT) encoders), whereas $\mathcal{K}_q^{gen}$ contains auto-regressive modules (e.g., LLMs) requiring prefill and iterative decoding.

		\item Dependencies ($\mathcal{E}_q$): The edges represent the data flow dependencies. An edge $(u, v) \in \mathcal{E}_q$ indicates that the output activations of module $u$ are required as input for module $v$.
	\end{itemize}

        \subsubsubsection{Fine-grained Layer Partitioning}
To support intra-module partitioning, each module $k \in \mathcal{K}_q$ is further partitioned into a sequence of layers $\mathcal{L}_k = \{1, \dots, L_k\}$. Note that any inter-module dependency $(u, v) \in \mathcal{E}_q$ is mathematically modeled as a direct connection from the final layer $L_u$ of module $u$ to the first layer of module $v$. For a layer $l$ of module $k$, we denote its computational workload as $w_{k,l}$, output activation tensor size as $\delta_{k,l}$ (transmitted if cut after layer $l$), static weight memory as $\mu_{k,l}^{stat}$, and task-varying dynamic memory (e.g., KV-cache) as $\mu_{q,k,l}^{dyn}$.

	\subsubsubsection{Decision Variables}
	EdgeCoInfer defines three decision variables for collaborative inference: $x_{q,k,l,n} \in \{0, 1\}$ indicates whether layer $l$ of module $k$ in task $q$ is assigned to node $n$; $\sigma_{k,l,n} \in \{0, 1\}$ indicates whether static weights of layer $l$ of module type $k$ are loaded on node $n$ to serve multiple tasks; and $f_{q,n}$ denotes the continuous operating frequency allocated to task $q$ on node $n$.


	\subsection{Computational Model}
	The computational process involves executing the specific layers of functional modules on edge nodes. We first quantify the workload based on the module type and then formulate the processing latency.
	
	\subsubsection{Workload Quantification}
	The computational workload $w_{k,l}$ depends on the specific architecture of layer $l$ in module $k$. We categorize modules into Standard Functional Modules and Generative Functional Modules.
	
	\textit{\quad a) Standard Functional Modules (e.g., Vision Encoders):} \begin{itemize}
		\item {CNN-based Layers:} For a convolutional layer with kernel size $K_l$, input channels $C_l^{in}$, and output channels $C_l^{out}$. We denote the output feature-map resolution of layer $l$ as $H_l^{out} \times W_l^{out}$. The workload is determined by the output feature map size \cite{sandler2018mobilenetv2}:
		\begin{equation}
			w_{k,l}^{cnn} = 2 H_l^{out}W_l^{out}  C_l^{in}  C_l^{out}  K_l^2,
		\end{equation}

		\item {ViT-based Layers:} For Vision Transformers, the sequence length is $s_{k,l} = H_l^{img} \cdot W_l^{img}/\rho^2$, where $H_l^{img}\times W_l^{img}$ is the input image resolution and $\rho$ is the patch size. The workload for a self-attention block is \cite{kaplan2020scaling}:
		\begin{equation}
			w_{k,l}^{vit} = 24 s_{k,l} h^2 + 4 {s_{k,l}}^2  h,
		\end{equation}
		where $h$ is the hidden dimension. It is worth noting that our framework can be readily extended to support other neural architectures beyond CNNs and ViTs by simply integrating their specific computational and memory profiling functions.
	\end{itemize}

	\textit{\quad b) Generative Functional Modules (e.g., LLM Decoders):} For generative modules, the workload varies dynamically with the sequence length. We distinguish between the prefill and decode phases.
	\begin{itemize}
		\item {Prefill Phase:} The entire prompt of sequence length $s_{k,l}^{in}$ is processed in parallel. The workload exhibits quadratic complexity due to the full self-attention mechanism \cite{korthikanti2023reducing}:
		\begin{equation}
			w_{k,l}^{pre} = 24  s_{k,l}^{in}  h^2 + 4 ({s_{k,l}^{in}})^2  h.
		\end{equation}
		\item {Decode Phase (with KV-Cache):} The model generates tokens auto-regressively and sequence length grows dynamically. Let $s_{k,l}^{out}$ denote the target or maximum number of tokens to be generated. For the $i$-th decoding step ($1 \le i \le s_{k,l}^{out}$), the effective context length includes the prompt and previously generated tokens $s_{k,l}[i] =s_{k,l}^{in} + i - 1$. Leveraging the KV-cache, the attention computation is reduced to linear complexity relative to the current context length:
		\begin{equation}
			w_{k,l}^{dec}(i) = 24  h^2 + 4  s_{k,l}[i]  h.
		\end{equation}
	\end{itemize}
	
	\subsubsection{Computational Latency and Energy}
	It is critical to distinguish between memory sharing and computation sharing. While static weights can be shared across tasks by storing a single copy, computation is inherently additive. Even if a functional module $k$ is reused by multiple tasks, the accelerator must execute the corresponding operations separately for their distinct inputs.
	
	We first denote $\psi_n$ as the effective throughput coefficient of node $n$ in FLOPs per Hz, which captures device-level effects such as parallelism, kernel efficiency, on-chip memory, memory bandwidth limitations and bandwidth contention. Let $f_{q,n}$ denote the effective accelerator operating frequency allocated to task $q$ on node $n$. This quantity can be interpreted as an equivalent value that jointly reflects Dynamic Voltage and Frequency Scaling (DVFS) and time-sliced scheduling on a shared accelerator. The effective compute throughput allocated to task $q$ on node $n$ is then $\psi_n f_{q,n}$ in FLOPs per second. Accordingly, the computation latency for executing layer $l$ of module $k$ for task $q$ on node $n$ is given by 
	\begin{equation}
		t_{q,k,l,n}^{comp} = \frac{w_{k,l}\,x_{q,k,l,n}}{\psi_n f_{q,n}} .
	\end{equation}
	
	We further model the computational power on on-chip accelerators using a DVFS-aware formulation consisting of a static leakage term and a frequency-dependent dynamic term. The computational power associated with task $q$ running on node $n$ is given by \cite{dso_gpu_dvfs}
	\begin{equation}
		P_{q,n}^{comp} = P_{n}^{sta} + \kappa_{n}\left(f_{q,n}\right)^{\alpha},
	\end{equation}
	where $P_{n}^{sta}$ is the static power, $\kappa_n$ is a device-dependent coefficient, and $\alpha$ is a hardware-dependent exponent.
	
	Therefore, the computational energy for executing layer $l$ of module $k$ for task $q$ on node $n$ is obtained as
	\begin{equation}
		E_{q,k,l,n}^{comp}
		= P_{q,n}^{comp}\, t_{q,k,l,n}^{comp}
		= \Big(P_{n}^{sta} + \kappa_{n}(f_{q,n})^{\alpha}\Big)\,
		\frac{w_{k,l}\,x_{q,k,l,n}}{\psi_n f_{q,n}} .
	\end{equation}

	\subsection{Memory Model}
	We model the memory footprint as the sum of static model parameters and dynamic runtime states.
	
	\subsubsection{Static Memory Modeling (Inter-Model Sharing)}The static memory consumption corresponds to the storage of model weights and biases. A key feature of EdgeCoInfer is the ability to share functional modules. The total static memory footprint on node $n$ is calculated based on the unique module layers loaded:
	\begin{equation}
		M_n^{stat} = \sum_{k \in \mathcal{K}} \sum_{l \in \mathcal{L}_k} \sigma_{k,l,n}  \mu_{k,l},
	\end{equation}
	The parameter size $\mu_{k,l}$ is determined by the specific layer architecture and data precision $b$ in bytes per element parameter. By adjusting $b$, the system can dynamically trade accuracy for memory/compute footprint, thereby adapting the runtime demand to the available device resources and maintaining model deployment feasibility.

	\begin{itemize} \item {CNN-based Layers:} For a convolutional layer, the parameters consist of the kernel weights and biases:
		 \begin{equation} 
			\mu_{k,l}^{cnn} = \left( C_l^{in}  C_l^{out}  K_l^2 + C_l^{out} \right)  b. 
	\end{equation}
		
	\item {Transformer-based Layers (ViT/LLM):} For a standard Transformer block, the parameters are concentrated in the Linear projections (Q, K, V, O) and the Feed-Forward Network (FFN). For a hidden dimension $h$, the parameter size is approximated as \cite{kaplan2020scaling}:
	\begin{equation}
		\mu_{k,l}^{trans} = 12 h^2  b.
	\end{equation}
	Note that different from computational workload, the static memory of a Transformer layer is independent of the sequence length.
\end{itemize}
	
	\subsubsection{Dynamic Memory Modeling (Intra-Module Partitioning)} Dynamic memory consists of intermediate activation tensors and, crucially for LLMs, the KV cache. This consumption is strictly task-specific and additive, meaning it cannot be shared across tasks.
	
	\subsubsubsection{Activation Memory}For standard layers, node $n$ must allocate buffer space to accommodate the activation tensor generated during the forward pass. Relating this to the output transmission size $\delta_{k,l}$, the required buffer space for layer $l$ is modeled as:
	\begin{equation}
		\mu_{k,l}^{act} = \eta_{k,l} \cdot \delta_{k,l},
	\end{equation}
	where $\eta_{k,l}$ is an architecture-dependent expansion factor, which is obtained via lightweight profiling.

	\subsubsubsection{KV-Cache Memory}For generative modules, the KV-cache grows linearly with the sequence length to support auto-regressive decoding. The reserved KV-cache size is formulated as \cite{su2025accurate}:
	\begin{equation}
		\mu_{q,k,l}^{kv} = 2 (s_{k,l}^{in} + s_{k,l}^{out} ) h  b.
	\end{equation}
	The constant 2 accounts for both Key and Value matrices.

    Since sequential execution reuses activation buffers, peak activation memory equals the largest assigned layer. Conversely, KV caches accumulate across all assigned layers. Thus, the total dynamic memory on node $n$ is:
	\begin{equation}
        M^{dyn}_n = \sum_{q \in Q} [ \max_{k \in \mathcal{K}_q, l \in \mathcal{L}_k} (x_{q,k,l,n} \mu^{act}_{q,k,l}) + \hspace{-3mm} \sum_{k \in \mathcal{K}_q^{gen}, l \in \mathcal{L}_k} \hspace{-3mm}x_{q,k,l,n} \mu^{kv}_{q,k,l} ].
        \label{eq:dynamic_memory}
        \end{equation}

	\subsection{Communication Model}
	The collaborative inference requires transmitting intermediate activation tensors between edge nodes when adjacent layers are partitioned across different devices. We first calculate the data size and then formulate the transmission latency.
	
	\subsubsection{Intermediate Data Size} 
	\subsubsubsection{Standard Functional Modules}The intermediate activation to be transmitted after layer $l$ equals the output feature map for CNNs or the token sequence for ViTs, and can be expressed as
	\begin{align} 
		\delta _{k,l}^{cnn} =	&{H_l^{out}}  {W_l^{out}}  C_l^{out}  b\\
	 \delta _{k,l}^{vit} =& {s_{k,l}}  h  b.
	\end{align}

	\subsubsubsection{Generative Functional Modules (LLM)}For LLMs, the communication overhead differs drastically between phases due to the auto-regressive nature.
	\begin{itemize}
		\item Prefill Phase: The output of layer $l$ contains the hidden states for all tokens in the prompt. For a sequence length $s$, the transmission volume is
		\begin{equation}
			\delta_{k,l}^{pre}(s) = s  h  b.
		\end{equation}
		\item Decode Phase: In each step of token generation, only the hidden state of the \textit{newest} token needs to be transmitted to the next layer (as previous keys/values are already cached at the destination). Thus, the transmission volume is minimal and constant, given by
		\begin{equation}
			\delta_{k,l}^{dec} = 1  h  b.
		\end{equation}
	\end{itemize}

	\subsubsection{Transmission Latency}
	Data transmission occurs strictly when layer $l$ and layer $l+1$ are assigned to different nodes. The total bandwidth $B$ is divided into $N$ orthogonal sub-channels to prevent interference. Therefore, the data transmission rate is calculated as:
	\begin{equation}\label{transmit}
		R_{n,m} = \frac{B}{N}    \log_2 \left( 1 + \frac{g_{n,m} P_n^{trans} }{\beta^2} \right),
	\end{equation}
	where $B$ is the bandwidth, $P_n^{trans}$ is the transmission power, $g_{n,m}$ represents the channel gain, and $\beta^2$ denotes the background Gaussian noise power.

	Therefore, the transmission latency for the output of layer $l$ (assigned to node $n$) to layer $l+1$ (assigned to node $m$) is
	\begin{equation}
		t_{q,k,l,n,m}^{trans} = \frac{8\delta_{k,l}  x_{q,k,l,n}  x_{q,k,l+1,m}}{R_{n,m}},
	\end{equation}
	where the constant $8$ converts the data size from bytes to bits. The product $x_{q,k,l,n}  x_{q,k,l+1,m}$ acts as a binary indicator that is $1$ only if the partition cut occurs exactly between layer $l$ and $l+1$. For the last layer $l$ of a module, $l+1$ denotes the first layer of the successor module connected through DAG edge.

	\subsubsection{Transmission Energy}
	The transmission energy is determined by the transmission power $P_n^{trans}$ and the duration of data transfer. Based on the transmission latency $t_{q,k,l,n,m}^{trans}$, the energy consumed by node $n$ to transmit the output of layer $l$ to node $m$ is given by
	\begin{equation}
		E_{q,k,l,n,m}^{trans} = P_n^{trans}  t_{q,k,l,n,m}^{trans} = \frac{ 8 P_n^{trans} \delta_{k,l}  x_{q,k,l,n}  x_{q,k,l+1,m}}{R_{n,m}}.
	\end{equation}
	Similar to latency, transmission energy is only incurred when adjacent layers are partitioned across different nodes ($n \ne m$).

	\subsection{Energy Consumption Model}
	Energy efficiency is paramount for battery-powered EDs. The total energy consumption must account for the repeated execution of the decoding phase. For any specific layer $l$ of module $k$ in task $q$, the total processing energy is the sum of its computational energy on the assigned node and the transmission energy to the subsequent node as follows:
	\begin{equation}
		E_{q,k,l} = \sum_{n \in \mathcal{N}}  ( E_{q,k,l,n}^{comp} + \sum_{m \in \mathcal{N} ,m\ne n}   E_{q,k,l,n,m}^{trans} )
	\end{equation}
	
	Task execution encompasses both the Standard Phase (executing non-generative modules) and the Generative Phase (comprising prefill and iterative decoding). Specifically, the total energy consumption for completing task $q$ is the aggregation of computation and transmission energy across all involved nodes and layers:
            \begin{align}
           & E_{q}^{total} =  \underbrace{\sum_{k \in \mathcal{K}_q^{std}, l \in \mathcal{L}_k} E_{q,k,l}(w_{k,l}, \delta_{k,l})}_{\text{Standard Phase (Non-generative)}} \\
            & + \underbrace{\sum_{k \in \mathcal{K}_q^{gen}, l \in \mathcal{L}_k} [ E_{q,k,l}(w_{k,l}^{pre}, \delta_{k,l}^{pre})}_{\text{Prefill Phase (One-time)}} + \underbrace{\sum_{i=1}^{s_{k,l}^{out}} E_{q,k,l}(w_{k,l}^{dec}(i), \delta_{k,l}^{dec}) ]}_{\text{Decode Phase (Iterative)}} \nonumber.
            \end{align}
	For standard tasks, the second and third terms naturally become zero, reducing the model to the classical DAG energy summation.

	\subsection{Inference Latency Model}
	The end-to-end inference latency is the primary metric for QoE. We formulate the total latency by aggregating the processing time across the task's specific execution phases: the one-time encoding/prefill and the iterative decoding. For any specific layer $l$ of module $k$, the processing delay $T_{q,k,l}$ consists of execution and transmission time. Crucially, the values depend on the specific phase given by
	\begin{equation}
		T_{q,k,l} = \sum_{n \in \mathcal{N}} ( t_{q,k,l,n}^{comp} + \sum_{m \in \mathcal{N},m\ne n} t_{q,k,l,n,m}^{trans} ).
	\end{equation}

	Since task $q$ is modeled as a DAG, standard modules may run in parallel branches. The total inference latency for completing task $q$ is the aggregation of computation and transmission delays across all involved nodes and layers given as follows:
        \begin{align}
       & L_{q}^{total} =  \underbrace{\max_{p \in \mathcal{P}_q^{std}} \sum_{(k,l) \in p} T_{q,k,l}(w_{k,l}, \delta_{k,l})}_{\text{Standard Path (Non-generative)}} \\
        & + \underbrace{\sum_{k \in \mathcal{K}_q^{gen}, l \in \mathcal{L}_k} [ T_{q,k,l}(w_{k,l}^{pre}, \delta_{k,l}^{pre})}_{\text{Prefill Phase (One-time)} } + \underbrace{\sum_{i=1}^{s_{k,l}^{out}} T_{q,k,l}(w_{k,l}^{dec}(i), \delta_{k,l}^{dec})}_{\text{Decode Phase (Iterative)}} ]\nonumber,
        \end{align}
	where $\mathcal{P}_q^{std}$ represents the set of paths in the subgraph formed by $\mathcal{K}_q^{std}$.

\subsection{Problem Formulation}


	
We introduce three sets of optimization variables $\{\mathbf{X}, \boldsymbol{\Sigma}, \mathbf{F}\}$ to orchestrate the collaborative inference. $\mathbf{X} = \{x_{q,k,l,n}\}$ denotes the partitioned layer decisions. $\boldsymbol{\Sigma} = \{\sigma_{k,l,n}\}$ represents the module sharing decisions. $\mathbf{F} = \{f_{q,n}\}$ denotes the allocation of computational resources.
	
	The system objective is to minimize the total system cost, defined as the weighted sum of energy consumption and inference latency across all concurrent tasks. This objective allows the system to balance execution speed and energy efficiency based on user preferences, which can be formulated as:
	\begin{align} 
		\label{equa: problem P1} 
		{\bf{P1}}:\mathop {\min }\limits_{{\bf{\Sigma }},{\bf{X}},{\bf{F}}} {\rm{ }}\sum\limits_{q \in {\cal Q}} [ {\lambda _q}  E_q^{total} + (1 - {\lambda _q})  \upsilon   L_q^{total}]
	\end{align}
	\textbf{Subject to} 
	\begin{align} 
		& \sum_{n \in \mathcal{N}} x_{q,k,l,n} = 1, \forall q \in \mathcal{Q}, k \in \mathcal{K}_q, l \in \mathcal{L}_k \label{const:assignment}\\ 
		& x_{q,k,l,n} \le \sigma_{k,l,n}, \forall q \in \mathcal{Q}, k \in \mathcal{K}_q, l \in \mathcal{L}_k, n \in \mathcal{N} \label{const:sharing_logic}\\
		&M_n^{stat} + M_n^{dyn} \le M_n^{max}, \quad \forall n \in \mathcal{N} \label{const:memory}\\
		& \sum_{q \in \mathcal{Q}} f_{q,n} \le f_n^{max}, \quad \forall n \in \mathcal{N} \label{const:compute}\\
		& L_q^{total} \le D_q, \quad \forall q \in \mathcal{Q} \label{const:deadline}
	\end{align}
	where $\lambda_q \in [0, 1]$ is the weighting factor balancing performance, and $\upsilon$ is a normalization coefficient to align the scales of time and energy. Constraint (\ref{const:assignment}) ensures structural integrity and guarantees that every layer $l$ of every module $k$ in task $q$ is assigned to exactly one compute node. Constraint (\ref{const:sharing_logic}) indicates that if layer $l$ of module type $k$ is used by any task $q$ on node $n$, the corresponding static weights must be loaded into memory. Constraint (\ref{const:memory}) ensures that the sum of unique static weights and additive dynamic KV-caches is constrained by the device budget $M_n^{max}$. Constraint (\ref{const:compute}) ensures that the allocated computational resources do not exceed the device capacity. Constraint (\ref{const:deadline}) imposes the QoS constraint, ensuring the latency satisfies the specific deadline $D_q$.

	\section{Hybrid Evolutionary Hierarchical Reinforcement Learning Framework}
	\label{sec:proposed_method}
	
	\subsection{Problem Analysis and Methodological Rationale}
	The formulated Problem \textbf{P1} constitutes a Mixed-Integer Non-Linear Programming (MINLP) challenge, which is NP-hard due to the combinatorial explosion of the joint discrete-continuous decision space and the severely restricted feasible region caused by coupled hard constraints on memory and latency. Conventional heuristics often are trapped in local optima, while standard DRL approaches struggle to converge due to the high dimensionality and sparsity of valid solutions. To tackle this intractability, we propose the HE-HRL framework, which introduces a methodological shift based on hierarchical decomposition and feasibility-guided learning.

	\begin{figure*}[t]
		\makebox[\textwidth][c]{\includegraphics[width=1\textwidth]{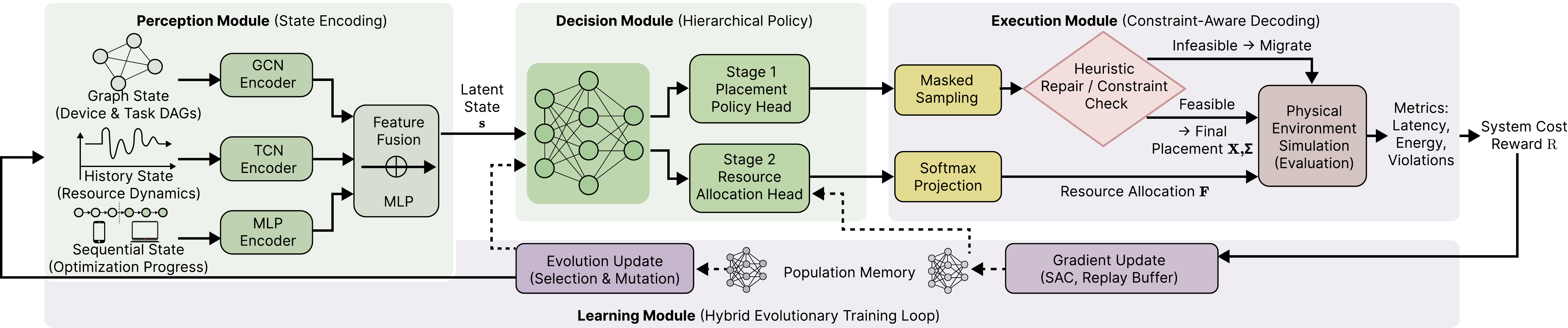}}
		\vspace{-2em}
		\caption{Architecture of the proposed HE-HRL framework. 
        }
		\label{fig:architecture}
	\end{figure*}	
	
	As shown in Fig. \ref{fig:architecture}, the architecture comprises four synergistic components where the Perception module extracts spatio-temporal features and the Decision module generates hierarchical policies. Subsequently, the Execution module ensures constraint satisfaction through pruning and repair, while the Learning module optimizes policies via a hybrid evolutionary-gradient paradigm. Details are provided below.

	\subsection{Perception Module: Spatio-Temporal State Encode}
	The Perception module constructs a comprehensive latent representation $\mathbf{s}$ from the system state. We design a hybrid encoder to capture three distinct views of the edge environment.
	
	\subsubsection{Graph State} To capture the topology-dependent resource heterogeneity, we represent the edge network infrastructure as a graph $\mathcal{G}=(\mathcal{N}, \mathcal{E})$, where $\mathcal{N}$ is the set of computing nodes and $\mathcal{E}$ represents network links between nodes. Each node $n$ possesses a feature vector $\mathbf{y}_n$, including memory budget, frequency, and current load, which can be expressed as:
	\begin{equation}
		\mathbf{y}_n = [ {M_n^{max}, f_n^{\max}, \psi_n}, {{M}_n, \sum\nolimits_q f_{q,n}} ]^T.
	\end{equation}
	
	We employ a Graph Convolutional Network (GCN) \cite{kipf2017semi} to aggregate neighborhood information via the adjacency matrix $\mathbf{A}$:
	\begin{equation}
		\mathbf{z}_{GCN} = \sigma\left( \hat{\mathbf{D}}^{-\frac{1}{2}} \hat{\mathbf{A}} \hat{\mathbf{D}}^{-\frac{1}{2}} \mathbf{Y} \mathbf{W}_G \right),
	\end{equation}
	where $\mathbf{Y}$ is the node feature matrix obtained by stacking all $\mathbf{y}_n$, $\hat{\mathbf{A}} = \mathbf{A} + \mathbf{I}$ incorporates self-loops, $\hat{\mathbf{D}}$ is the degree matrix of $\hat{\mathbf{A}}$, and $\mathbf{W}_G$ denotes the learnable weight matrix.
	
	\subsubsection{History State} Since edge dynamics exhibit strong temporal correlations, we encode the sliding window history $\mathbf{H}_t = [\mathbf{Y}_{t-K+1}, \dots, \mathbf{Y}_t]$ using a Temporal Convolutional Network (TCN) \cite{bai2018empirical}, where $K$ is the length of the historical window. The TCN employs dilated causal convolutions to expand the receptive field, extracting a trend embedding $\mathbf{z}_{TCN}$.

	\subsubsection{Sequential State} Since the partitioned and sharing strategy adopts a constructive cut-step mechanism (detailed in Section \ref{sec.Decision Module}), the decision process is inherently sequential. To ensure the state representation satisfies the Markov property, the agent requires explicit awareness of the current optimization progress. We encode the model partitioned status as a progress vector $\mathbf{p}_t$ as follows:
	\begin{equation}
		\mathbf{p}_t = [ {k}, {l}, L_{\text{part}},{L_{\text{rem}}} ]^T,
	\end{equation}
	where $k$ denotes the topological index of the current functional module, $l$ indicates the starting layer of the current partition, $L_{\text{part}}$ is the length of the current partition, and $L_{\text{rem}}$ represents the number of remaining layers to be assigned. This vector is projected into a dense embedding $\mathbf{z}_{\text{prog}}$ via a Multi-Layer Perceptron (MLP), providing the necessary temporal context for the sequential decision-making.
	
	\subsubsection{Feature Fusion} The final latent state $\mathbf{s}$ is obtained by concatenating and projecting these embeddings:
	\begin{equation}
		\mathbf{s} = \text{MLP}( \mathbf{z}_{\text{GCN}} \mathbin{||} \mathbf{z}_{\text{TCN}} \mathbin{||} \mathbf{z}_{\text{prog}} ).
	\end{equation}

	\subsection{Decision Module: Hierarchical Policy Network} \label{sec.Decision Module}
	The decision module serves as the core mapping engine, transforming the latent state $\mathbf{s}$ into the complete set of optimization variables $\{\mathbf{X}, \boldsymbol{\Sigma}, \mathbf{F}\}$ defined in Problem \textbf{P1}. We employ a two-stage neural architecture $\pi_{\theta}(\mathbf{a}|\mathbf{s})$ to decouple the discrete partitioned and sharing decisions from the continuous resource allocation.
	
	\subsubsection{Partitioned and Sharing Policy of Stage-1}
	The high-level policy $\pi_{\theta_1}(\cdot|\mathbf{s})$ constructs the binary partitioned variables $\mathbf{X}$ via a sequential Cut-Step mechanism. Instead of outputting the high-dimensional decision matrix directly, it iteratively generates partitions. At each step, the policy network produces two sets of unnormalized scores, denoted as logits $\mathbf{L}$. First is Cut Logits $\mathbf{L}_{\text{cut}} \in \mathbb{R}^{L_{\text{rem}}}$, which is a vector of scores over the layer slots. These logits determine the partition length $\ell$ (i.e., how many consecutive layers to group) by selecting the cut position with the highest score. Second is Node Logits $\mathbf{L}_{\text{node}} \in \mathbb{R}^{N}$, a vector of scores over the $N$ available computing nodes. These logits identify the destination node $n$ for the current partition via a sampling operation.

    In addition, this stage determines the module sharing variables $\boldsymbol{\Sigma}$ to implement memory deduplication. Given the partitioning decision $\boldsymbol{X}$, we deterministically derive $\sigma_{k,l,n}=\bigvee_{q\in\mathcal{Q}}x_{q,k,l,n}$, where $\bigvee$ denotes the logical OR operation. This mapping ensures that the static weights of layer $l$ in module $k$ are instantiated only once on node $n$, even when requested by multiple concurrent tasks. It also satisfies the sharing constraint by construction while reducing redundant exploration in the discrete action space.

	\subsubsection{Resource Allocation Policy of Stage-2}
	The low-level policy $\pi_{\theta_2}(\cdot|\mathbf{s})$ governs the allocation of resources to satisfy the system demands established by Stage-1. To facilitate stable training under strict constraints, we design a deterministic controller to provide feasible feedback for the high-level Stage-1 policy in the early training stage. Once Stage-1 stabilizes, we activate the neural network to perform fine-grained optimization. Specifically, regarding computational resources ($\mathbf{F}$), the policy outputs a continuous score matrix $\mathbf{L}_{\text{freq}} \in \mathbb{R}^{N \times Q}$, where $N$ is the number of nodes and $Q$ is the number of concurrent tasks.

	\subsection{Execution Module: Constraint-Aware Action Decoding}
	Directly sampling from the policy network often leads to invalid actions in constrained edge environments \cite{9530374}. This module implements the Feasibility-Guided Decoding mechanism. 
	
	\subsubsection{Feasibility Pruning and Heuristic Repair for Stage-1}
	We treat the partitioned and sharing decision generation as a constructive process. To ensure feasibility, we integrate constraint checks directly into the decoding loop.
	
	First is the pre-action pruning. Before committing a candidate partition of length $\ell$ to node $n$, we compute a validity mask $\mathcal{M}(\ell, n)$. We define the memory feasibility condition as:
		\begin{equation}
			M_n + \Delta M(\ell) \le M_n^{max},
		\end{equation}
		where $M_n$ denotes the current memory occupancy of node $n$, calculated as the sum of all layers previously assigned to this node. $\Delta M(\ell)$ represents the incremental memory cost of the new candidate partition. Moreover, the estimated cumulative latency must not exceed the task deadline $D_q$.
	\begin{equation}
		T_q^{\text{ela}} + T_q^{\text{est}}(\ell, n) \le D_q,
	\end{equation}
	where $T_q^{\text{ela}}$ denotes the elapsed time, and $T_q^{\text{est}}(\ell, n)$ is a Stage-2-independent optimistic lower bound computed using $f_n^{\max}$ and zero additional contention, so only candidates that cannot meet $D_q$ under any feasible resource allocation are pruned. The mask $\mathcal{M}(\ell, n)$ is set to $-\infty$ if this condition is violated or if the constraint bound is breached. The policy distribution is then re-normalized:
	\begin{equation}
		P(\ell, n | \mathbf{s}) = \text{Softmax}\left( (\mathbf{L}_{\text{cut}} \oplus \mathbf{L}_{\text{node}}) + \mathcal{M} \right),
	\end{equation}
	where $\oplus$ denotes the broadcasting operation, resulting in a matrix of size $S \times N$. 
		
	Second is the post-action repair. If an action still violates deadlines or causes Out-Of-Memory errors during iteration, a heuristic repair operator is triggered. It iteratively migrates bottleneck partitions to the MEC server or nodes with maximal resource slack.
	
	\subsubsection{Deterministic Controller and Differentiable Projection for Stage-2}
	To strictly satisfy the capacity constraint, we employ a two-phase curriculum strategy that transitions from a heuristic controller to a differentiable projection.
	
	During the initial training phase, we employ a feasibility-preserving heuristic to compute $\mathbf{F}$ directly from $(\mathbf{X},\boldsymbol{\Sigma})$. We adopt a Deadline-Aware Load Balancing rule. The frequency $f_{q,n}$ is allocated proportional to the urgency given as
	\begin{equation}
		f_{q,n} = f_n^{\max} \cdot \frac{a_{q,n}}{\sum_{i=1}^{Q} a_{i,n}},  a_{q,n} = \frac{w_{q,n}}{D_q}.
	\end{equation}
	where $w_{q,n}$ denotes the workload of task $q$ placed on node $n$.
	
	In the refinement phase, we replace softmax with a masked bounded-sigmoid projection to enable energy-saving frequency downscaling. Let $l^{(f)}_{n,q}$ be the unnormalized logit and $m_{q,n} = \bigvee_{k,l} x_{q,k,l,n}$ be the binary placement mask. The operating frequency is:
\begin{equation}
f_{q,n} = f_n^{max} \cdot \frac{\varphi(l^{(f)}_{n,q}) \cdot m_{q,n}}{\max \left( 1, \sum_{i=1}^Q \varphi(l^{(f)}_{n,i}) m_{i,n} \right)},
\label{eq:freq_allocation}
\end{equation}
where $\varphi(\cdot)$ is the sigmoid activation function. This strictly enforces $f_{q,n} = 0$ for absent tasks and $\sum_{q} f_{q,n} \le f_n^{max}$, while permitting the total frequency to drop below $f_n^{max}$ to save energy via DVFS.

	Consequently, we obtain the complete set of decision variables $(\mathbf{X}, \boldsymbol{\Sigma}, \mathbf{F})$, from which the total system cost is derived. The final reward $R$ is defined as the negative of the total system cost.
	
	\subsection{Learning Module: Hybrid Evolutionary Training}
	Since the decision variables contain both discrete and continuous, the overall optimization problem is non-differentiable. We decouple the training into two synchronized paradigms.

	\subsubsection{Evolutionary Path for Stage-1 Optimization}
	The high-level policy $\pi_{\theta_1}$ generates partitioned and sharing logits, which are mapped to discrete actions $(\mathbf{X},\boldsymbol{\Sigma})$ via the pruning and sampling process. Since gradients cannot backpropagate through these discrete and heuristic operations, we employ a gradient-free GA.

	We maintain a population of $P$ actors $\mathcal{P} = \{\theta_1^1, \dots, \theta_1^P\}$. In each generation $g$, the fitness of actor $i$ is evaluated as its average episodic return $J(\theta_1^i) = \mathbb{E}[R]$.

	For the selection operator in GA, we rank actors by fitness and select the top-$k$ elites, denoted as $\mathcal{P}_{\text{elite}} \subset \mathcal{P}$.
	
	For the mutation operator in GA we use Parameter Space Noise. To explore the combinatorial solution space, we generate an offspring parameter $\tilde{\theta}_1$ for each elite $\theta_1 \in \mathcal{P}_{\text{elite}}$ by injecting random perturbations:
	\begin{equation}
		\tilde{\theta}_1 = \theta_1 + {\epsilon},  {\epsilon} \sim \mathcal{N}(\mathbf{0}, \vartheta^2 \mathbf{I}),
	\end{equation}
	where ${\epsilon}$ is the Gaussian noise vector, $\vartheta$ is the mutation strength, and $\mathbf{I}$ is the identity matrix. This zero-order optimization allows the agent to escape local optima in the rugged discrete landscape. 
	
	\subsubsection{Gradient Path for Stage-2 Optimization}
	The low-level policy $\pi_{\theta_2}$ generates the resource allocation. We train $\theta_2$ via SAC \cite{haarnoja2018soft} to maximize the entropy-regularized objective:
	\begin{equation}
		J(\theta_2) = \mathbb{E}_{\tau \sim \pi_{\theta_2}} \left[ \sum_{t=0}^{T} \gamma^t \left( r(\mathbf{s}_t, \mathbf{a}_{t}) + \varsigma \mathcal{H}(\pi_{\theta_2}(\cdot|\mathbf{s}_t)) \right) \right],
	\end{equation}
	where $\tau = \{ (\mathbf{s}_t, \mathbf{a}_{t}) \}_{t=0}^{T}$ denotes the trajectory induced by the policy over a finite horizon $T$, $\gamma$ is the discount factor, and $\varsigma$ determines the relative importance of the entropy term $\mathcal{H}(\cdot)$.
	
	To optimize this objective, we use transitions $(\mathbf{s}, \mathbf{a}, r, \mathbf{s}')$ stored in replay buffer $\mathcal{B}$. Therefore, SAC optimization alternates between evaluating the policy via soft Q-value estimation (Critic) and improving the policy via gradient ascent (Actor). 
	
	First is the Critic Update. We maintain two Q-functions parameterized by $\phi_{a,b}$ to mitigate overestimation. The parameters are updated by minimizing the Bellman residual:
	\begin{equation}
		\mathcal{L}_Q(\phi_i) = \mathbb{E}_{\mathcal{B}} \left[ (Q_{\phi_i}(\mathbf{s}, \mathbf{a}) - y)^2 \right], \text{for } i \in \{a, b\},
	\end{equation}
	where the target $y$ is calculated using the target networks $\bar{\phi}$ and $\bar{\theta}_2$:
	\begin{equation}
		y = r + \gamma \min_{j=a,b} Q_{\bar{\phi}_j}(\mathbf{s}', \mathbf{a}') - \varsigma \log \pi_{{\theta}_2}(\mathbf{a}'|\mathbf{s}').
	\end{equation}

	Second is the Actor Update. The policy parameters $\theta_2$ are updated by minimizing the Kullback-Leibler (KL) divergence. Moreover, the reparameterization trick $\mathbf{a} = f_{\theta_2}(\mathbf{s}, \xi), \xi \sim \mathcal{N}(\mathbf{0}, \mathbf{I})$ is employed to allow gradients to propagate through the stochastic sampling process. Thus, the loss function can be expressed as:
\vspace{-2mm}
	\begin{equation}
		\begin{split}
			\mathcal{L}_\pi(\theta_2) = \mathbb{E}_{\mathbf{s} \sim \mathcal{B}, \xi \sim \mathcal{N}} \bigg[ & \varsigma \log \pi_{\theta_2}(f_{\theta_2}(\mathbf{s}, \xi)|\mathbf{s}) \\
			& - \min_{j=a,b} Q_{\phi_j}(\mathbf{s}, f_{\theta_2}(\mathbf{s}, \xi)) \bigg].
		\end{split}
	\end{equation}
	
	Once trained and deployed, the learned policy functions as a lightweight, real-time decision engine. In contrast to traditional iterative solvers that suffer from high computational latency, our approach directly maps real-time graph states to near-optimal partitioning and sharing decisions in short time. Crucially, thanks to the GCN-based structural perception, the algorithm generalizes well to unseen task topologies and fluctuating network conditions, enabling edge clusters to dynamically adapt to different workloads to achieve on-device MLLMs inference.
	
\subsection{Computational Complexity Analysis}
We analyze the computational complexity of EdgeCoInfer's inference phase.
At each Cut-Step action, the GCN and TCN encoders incur
$O(L_g(|\mathcal{E}|d+|\mathcal{N}|d^2))$ and
$O(|\mathcal{N}|Kd^2)$, respectively. Let $S$ be the total number of
layer slots and $A\leq S$ the number of Cut-Step actions. At step $t$,
the Stage-1 policy incurs
$O(H^2+H(S_t^{\rm rem}+|\mathcal{N}|)
+S_t^{\rm rem}\log S_t^{\rm rem}
+|\mathcal{N}|\log|\mathcal{N}|
+K_{\rm cut}K_{\rm node}\bar{S}_t)$,
where $K_{\rm cut}$ and $K_{\rm node}$ are fixed top-$K$
hyperparameters and $\bar{S}_t$ denotes the maximum number of layers
examined during a feasibility check. The total Stage-1 complexity is
the sum of the encoding and decoding costs over $A$ actions. In the
worst case, $\sum_t S_t^{\rm rem}=O(S^2)$, and hence the complexity
remains polynomial in $S$ and $|\mathcal{N}|$. Stage-2 is executed once with complexity
$O(H^2+H|\mathcal{N}|Q)$, while the masked frequency projection costs
$O(|\mathcal{N}|Q)$. Therefore, the overall online decision complexity
is polynomial in $S$, $Q$, and $|\mathcal{N}|$, avoiding exhaustive
enumeration of the combinatorial search space.

	\section{Experimental Study}\label{sec:Experimental Study}
	\subsubsection{Experimental Settings}
	Our simulation constructs a real-world edge computing environment consisting of varying numbers of nodes $N = 5, 10, \dots, 50$, utilizing empirical profiling data from multimodal models. Task generation follows a Poisson process. The detailed specifications of all functional modules are provided in Table~\ref{tab:component_specs}. We evaluate our system on six types of multimodal AI tasks as shown in Table~\ref{tab:task_modules}, which summarizes the module compositions for each task type. The network architectures and algorithmic hyperparameters are primarily adapted from \cite{haarnoja2018soft, 10623529, kipf2017semi}. The detailed configurations are summarized in Table \ref{tab:exp_settings}.

	\begin{table}[t]
		\vspace{-0.5em}
		\scriptsize
		\centering
		\caption{Module Specifications}
		\vspace{-1em}
		\label{tab:component_specs}
		\begin{tabular}{llcr}
			\toprule
			Functional Module& Model  & Params &Precision  \\
			\midrule
			Audio Encoder
        & ViT-B \cite{wu2023large}
        & 84.93 M & FP16 \\

        LLM Decoder
        & Vicuna-7B \cite{vicuna2023}
        & 6.44 B & FP16 \\

        Text Encoder
        & OpenCLIP TRF \cite{ilharco2021openclip}
        & 84.93 M & FP16 \\

        Visual Encoder
        & ViT-L/14 \cite{radford2021learning}
        & 301.99 M & FP16 \\

        Classification Head
        & Linear Classifier
        & 1.024 M & FP16 \\

        Alignment Head
        & Cosine Similarity
        & -- & FP16 \\
			\bottomrule
		\end{tabular}
	\end{table}
	
	\begin{table}[t]
		\centering
		\scriptsize
		\caption{Task Types and Module Compositions}
		\vspace{-1em}
		\renewcommand\arraystretch{1}
		\label{tab:task_modules}
		\setlength{\tabcolsep}{0.2mm}{
			\begin{tabular}{lp{5cm}}
				\toprule
				Task Type & Module Compositions \\
				\midrule
				Image Classification & ViT-L/14 $\to$ Linear Classifier \\
				Visual Question Answering & ViT-L/14 $\to$ Vicuna-7B \\
				Image Captioning & ViT-L/14 $\to$ Vicuna-7B \\
				Automatic Speech Recognition (ASR) & ViT-B $\to$ Vicuna-7B \\
				Audio-Visual Alignment & ViT-L/14 + ViT-B $\to$ Cosine Similarity\\
				Image-Text Retrieval & ViT-L/14 + OpenCLIP TRF $\to$ Cosine Similarity \\
				\bottomrule
		\end{tabular}}
	\end{table}
    
	\begin{table}[t]
            \vspace{-0.5em}
		\caption{Detailed Experimental Parameters}
            \vspace{-0.8em}
		\label{tab:exp_settings}
		\centering
		\resizebox{\linewidth}{!}{%
			\begin{tabular}{ll}
				\toprule
				\textbf{Parameter} & \textbf{Value} \\
				\midrule
				\multicolumn{2}{c}{\textit{Device \& Network Settings}} \\
				\midrule
				MEC Server ($f_0^{max}$, $M_0^{max}$, $\psi_0 f_0^{max}$) & 1.3 GHz, 16 GiB, 10.6 TFLOPS \\
				MEC Server Power ($P_0^{sta}$, $P_0^{trans}$) & 15 W, 2 W \\
				Edge Devices ($f_n^{max}$, $M_n^{max}$, $\psi_n f_n^{max}$) & $U(0.3, 1.2)$ GHz, $\{4,6,8\}$ GiB, 2.0 TFLOPS \\
				Edge Devices Power ($P_n^{sta}$, $P_n^{trans}$) & $U(0.8, 3.0)$ W, $U(0.8, 2.0)$ W \\
				DVFS Coefficients ($\kappa_n$, $\alpha$) & $10^{-27}$, 3.0 \\
				Communication Bandwidth ($B$), Noise ($\beta^2$) & 30 MHz, -113 dBm \\
				\midrule
				\multicolumn{2}{c}{\textit{Task \& Model Settings}} \\
				\midrule
				Parameter Precision ($b$) & FP16 (2 bytes) \\
				Task Deadlines $D_q$ (Standard / LLM) & 3.0 s / 8.0 s \\
				Cost Objective Weights ($\lambda_q$, $\upsilon$) & 0.6, 1.0 \\
				\midrule
				\multicolumn{2}{c}{\textit{HE-HRL Algorithm Hyperparameters}} \\
				\midrule
				Neural Dims (Embed, GCN/TCN, SAC) & 128, 128, 256 \\
				TCN Setup (Levels, Kernel, Window $K$) & 3, 3, 8 \\
				Evolution Params (Pop $P$, Elite $k$, Mut $\vartheta$) & 64, 4, 0.02 \\
				SAC Buffer Size, Batch Size, Warmup & 200,000, 128, 128 \\
				SAC Learning Rate, Updates per iter. & $3 \times 10^{-4}$, 16 \\
				SAC $\tau$, Entropy Coeff. ($\varsigma$) & 0.005, 0.2 \\
				\bottomrule
			\end{tabular}%
		}
	\end{table}

    The effectiveness of EdgeCoInfer is evaluated using the metrics of system cost and inference task completion rate. Consistent with existing distributed inference literature~\cite{mohammed2020distributed, li2024distributed, zhang2024edgeshard}, our evaluation focuses on system-level performance (latency, energy, and memory) rather than task accuracy. Under uniform floating-point precision without model quantization or structural modification, layer partitioning and sharing maintain mathematical equivalence to monolithic execution.

	\subsubsection{Comparison Methods}
	We compare our proposed method against the following methods:
	\begin{itemize}
		\item Baselines: SOTA baselines include {S2M3}~\cite{yoon2025s2m3}, which performs module sharing without layer partitioning, and {PCI}~\cite{11183748}, which partitions model layers across devices without module sharing; conventional baselines include {Full Local}, all inference tasks are performed locally; {Full MEC}, all tasks attempt to be offloaded to MEC; {Greedy}, where all tasks offload to the node with the most abundant resources.
		\item RL Comparisons: {TD3} \cite{10623529}, a twin-delayed deep deterministic policy gradient algorithm for continuous control; {PPO} \cite{schulman2017proximal}, a proximal policy optimization algorithm with clipped policy updates. All methods use the same parameter settings as ours.
		\item Ablations: {DP Scheduling}, where using dynamic programming decoding replaces cut step in Stage-1; {No Pruning}, which disables Stage-1 feasibility-guided pruning; {No Controller}, which disables Stage-2 load-proportional controller.
	\end{itemize}

        \begin{figure*}[htbp] 
		\centering
		
		\begin{minipage}[b]{0.68\textwidth}
			\centering
			\includegraphics[width=\linewidth]{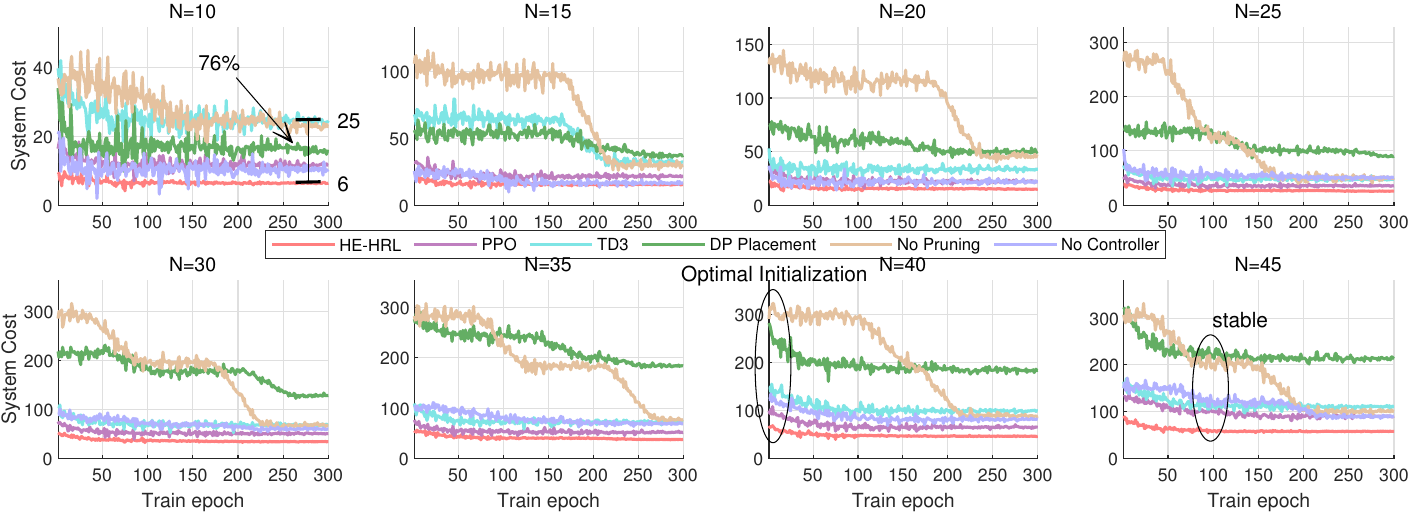} 
                \vspace{-8mm} 
			\caption{Convergence performance comparison.}
			\label{fig:convergence}
		\end{minipage}
		\begin{minipage}[b]{0.31\textwidth}
			\centering
			
			\includegraphics[width=\linewidth]{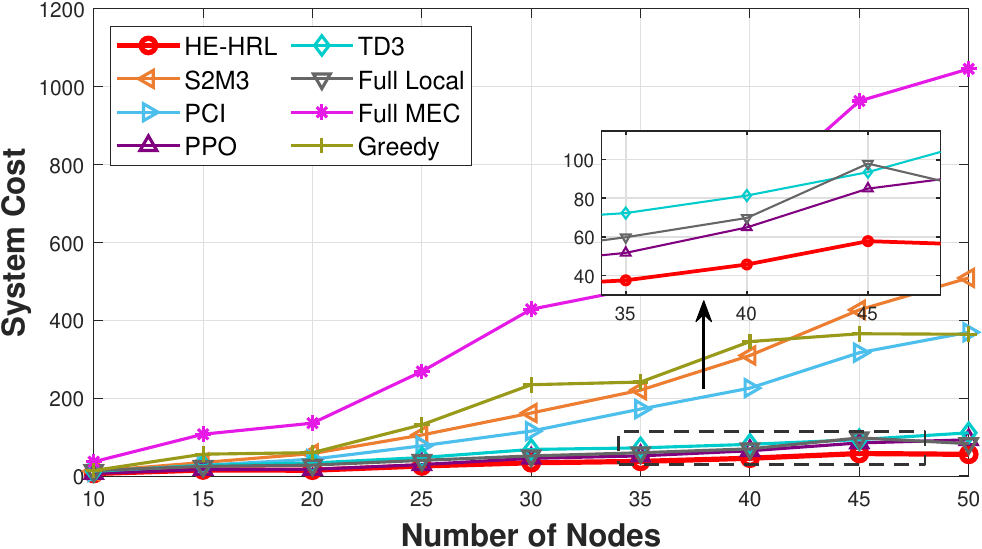} 
			
			\includegraphics[width=\linewidth]{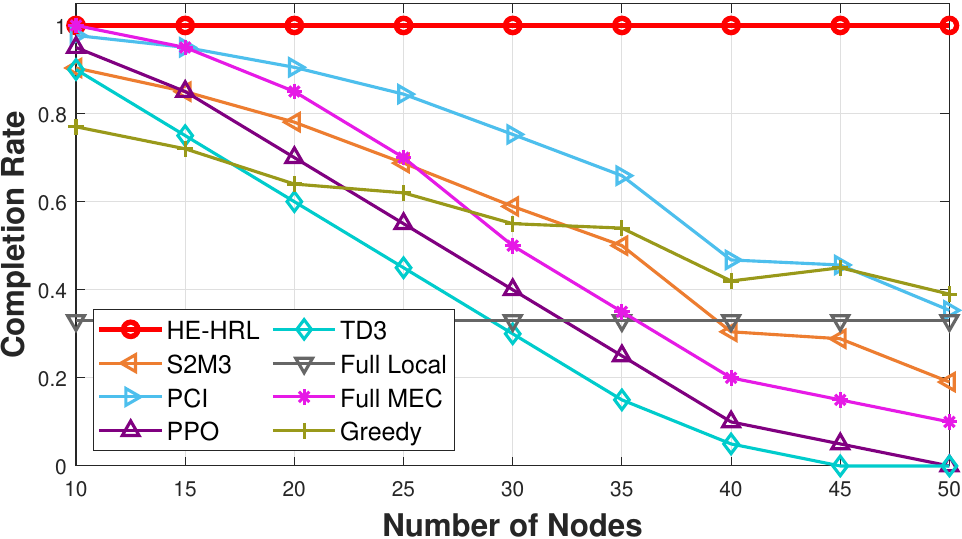} 
			\vspace{-8mm} 
			\caption{System cost and task completion rate comparison.}
			\label{fig:cost_and_completion}
		\end{minipage}
		
		\vspace{-2mm} 
	\end{figure*}


	\subsubsection{Convergence Analysis}
	
	Figure~\ref{fig:convergence} presents training convergence comparison, showing that HE-HRL achieves the lowest system cost with rapid and stable convergence. For instance, HE-HRL achieves a 76\% cost reduction at $N=10$. PPO and TD3 exhibit oscillatory curves with cost spikes, as uniform exploration wastes budget on infeasible configurations violating constraints. No Pruning consistently shows the highest costs and very slow convergence, particularly at larger $N$, with costs starting around 300 and declining only gradually—this unequivocally demonstrates that feasibility-guided pruning is essential for efficient search space exploration. DP Scheduling shows gradual cost reduction but plateaus at much higher levels, indicating that DP alone lacks adaptive optimization capabilities and cannot dynamically adjust to changing system loads. No Controller performs closer to PPO/TD3 but still at higher costs, showing that the load-proportional controller contributes to fine-tuning resource allocation in Stage-2. 

	
    \subsubsection{System Cost Analysis}

    The upper plot of Fig.~\ref{fig:cost_and_completion} shows that HE-HRL consistently achieves the lowest system cost across all evaluated scales, with the performance gap widening as the system scale increases. Across all evaluated scales and comparison methods, HE-HRL reduces system cost by an average of 59.86\%. The baseline methods exhibit fundamental limitations. Full MEC incurs the highest cost due to the massive data transmission volumes over constrained wireless channels and the severe compute bottleneck at the MEC server; Greedy ignores global resource optimization and energy-latency trade-offs; Full Local shows higher costs as local EDs lack sufficient computational capacity for LLM-based tasks. The RL algorithms PPO and TD3 achieve better performance than conventional heuristics but consistently fall short of HE-HRL, as the vast state-action space and lack of feasibility guidance make exploration challenging for general-purpose RL. S2M3 incurs rapidly escalating costs because complete modules cannot be split across devices, concentrating large LLM modules on a small number of high-memory nodes and causing severe memory contention; PCI outperforms S2M3 by partitioning model layers across nodes, yet still yields significantly higher costs than HE-HRL due to redundant memory overhead from independent static weight copies across concurrent tasks. In contrast, HE-HRL eliminates these memory and compute bottlenecks through joint model partitioning and weight sharing.
	
	
	\subsubsection{Task Completion Rate Analysis}
        The lower plot of Fig.~\ref{fig:cost_and_completion} demonstrates that HE-HRL achieves a perfect 100\% completion rate across all scales, while all other methods experience significant degradation as the system scales. Both PPO and TD3 start with high completion rates at small scales but decline rapidly as the number of nodes grows, revealing that general-purpose RL struggles with the problem's complexity and large state-action space. PPO slightly outperforms TD3 owing to its on-policy nature and clipped objective function, which provide more stable policy updates in discrete action spaces. For the ablation baselines, S2M3 experiences a sharp drop in completion rate due to single-device capacity bottlenecks, whereas PCI achieves higher completion rates than S2M3 but still degrades to 36\% under high concurrency. These results highlight the necessity of joint model partitioning and weight sharing, which enables HE-HRL to maintain robust service guarantees under dense task loads.
    
	
	\begin{table}[t]
		\centering
		\setlength{\tabcolsep}{3pt} 
		\renewcommand{\arraystretch}{1.1} 
		\caption{Static Memory Optimization Analysis: System Total Requirements vs. Optimized Loaded Memory ($N=40$)}
		\vspace{-1em}
		\label{tab:memory_optimization}
		\resizebox{\columnwidth}{!}{%
			\begin{tabular}{l c c r c c c r r}
				\toprule
				\multirow{2}{*}{\textbf{Model Instance}} & \multirow{2}{*}{\textbf{\shortstack{Unit\\Mem}}} & \multicolumn{2}{c}{\textbf{Requirement}} & & \multicolumn{3}{c}{\textbf{EdgeCoInfer Actual Loaded}} & \multirow{2}{*}{\textbf{\shortstack{Sav.\\(\%)}}} \\
				\cmidrule{3-4} \cmidrule{6-8}
				& & \textbf{Qty} & \textbf{Mem (GB)} & & \textbf{Qty} & \textbf{Partitioned} & \textbf{Mem (GB)} & \\
				\midrule
                ViT-L/14
                & 603.98 MB & 27 & 16.31 & & 3 & 14 & 1.81 & 88.89 \\
                Vicuna-7B
                & 12.89 GB & 10 & 128.85 & & 3 & 19 & 38.66 & 70.00 \\
                OpenCLIP TRF
                & 169.87 MB & 6 & 1.02 & & 2 & 2 & 0.34 & 66.67 \\
                ViT-B
                & 169.87 MB & 5 & 0.85 & & 2 & 2 & 0.34 & 60.00 \\
                Linear Classifier
                & 2.05 MB & 6 & 0.012 & & 1 & 0 & 0.002 & 83.33 \\
                Cosine Similarity
                & -- & 11 & -- & & 1 & 0 & -- & -- \\
                \midrule
                \textbf{Total}
                & \textbf{--} & \textbf{--} & \textbf{147.04}
                & & \textbf{--} & \textbf{--} & \textbf{41.15}
                & \textbf{72.02} \\
				\bottomrule
			\end{tabular}%
		}
	\end{table}

	\begin{table}[t]
		\centering
		\setlength{\tabcolsep}{4pt}
		\renewcommand{\arraystretch}{1.1}
		\caption{Dynamic Memory Optimization Analysis: Memory Composition of the Highest-Demand Device ($N=50$)}
		\vspace{-0.8em}
		\label{tab:dynamic_memory}
		\resizebox{\columnwidth}{!}{%
		\begin{tabular}{lcccccc}
			\toprule
			\textbf{Method} & \textbf{Static} & \textbf{Activation} & \textbf{KV Cache} & \textbf{Dynamic} & \textbf{Total Demand} & \textbf{Max. Excess} \\
			\midrule
        Full MEC
        & 13.83 & 0.22 & 7.25 & 7.47 & 21.30 & 4.12 \\

        S2M3
        & 13.49 & 0.10 & 2.42 & 2.52 & 16.01 & 4.62 \\

        PCI
        & 13.20 & 0.04 & 0.72 & 0.76 & 13.96 & 3.85 \\

        EdgeCoInfer
        & 7.22 & 0.02 & 0.46 & 0.48 & 7.70 & 0.00 \\
        \midrule
        \textbf{Avg. Reduction}
        & {46.53\%}
        & {73.64\%}
        & {70.25\%}
        & {70.46\%}
        & \textbf{53.53\%}
        & \textbf{--} \\
			\bottomrule
		\end{tabular}}
	\end{table}

	\subsubsection{Memory Optimization Analysis}
	Tables~\ref{tab:memory_optimization} and~\ref{tab:dynamic_memory} evaluate the static and dynamic memory efficiency of EdgeCoInfer. Table~\ref{tab:memory_optimization} compares the system total memory required by loaded model instances with the optimized actual static footprint under EdgeCoInfer. Cross-task module sharing reduces the number of loaded instances from 65 to 12, while layer partitioning distributes large models across devices without complete per-device replicas. For example, ViT-L/14 memory decreases from 16.31~GB to 1.81~GB, while Vicuna-7B achieves a 70\% reduction. Overall, EdgeCoInfer reduces static memory from 147.04~GB to 41.15~GB, saving 72.02\%.

	Table~\ref{tab:dynamic_memory} decomposes the peak device memory demand at $N=50$. Even under OOM conditions, baselines are evaluated by quantifying memory overflow as capacity excess. Full MEC concentrates workloads on the server, generating a 7.25~GB KV cache and a peak demand of 21.30~GB. S2M3 reduces the dynamic footprint but cannot partition complete modules across devices, resulting in substantial static memory demand and a 4.62~GB capacity excess. PCI distributes model layers and further reduces dynamic memory, but its lack of cross-task module sharing retains redundant model weights, leading to 13.20~GB of static memory and a 3.85~GB capacity excess. In contrast, EdgeCoInfer limits static, activation, and KV-cache memory to 7.22~GB, 0.02~GB, and 0.46~GB, achieving 7.70~GB total demand with zero capacity excess. On average, EdgeCoInfer reduces static, dynamic, and total memory demand by 46.53\%, 70.46\%, and 53.53\%, respectively.

	\subsubsection{Computational Efficiency Analysis}
	Beyond optimization performance, the decision-making latency is critical for real-time edge scheduling. We evaluated the average inference time of the trained HE-HRL policy on the MEC server. The results show that our framework generates a valid joint strategy in only 15.17 ms on average. Compared to the task execution deadlines, this decision overhead is negligible. This confirms that EdgeCoInfer supports online scheduling with millisecond-level responsiveness.
	\section{Conclusion}\label{sec:Conclusion}
	This paper proposed EdgeCoInfer, a hierarchical collaborative inference framework for concurrent MLLM execution on edge devices. EdgeCoInfer enables coarse-grained module sharing and fine-grained layer partitioning to overcome memory and concurrency bottlenecks. We tackle the non-differentiable joint optimization problem via a novel HE-HRL framework. Extensive evaluations validate that EdgeCoInfer effectively breaks the edge memory wall, reducing memory consumption by 53.53\% and system cost by 59.86\% over existing methods.
	

	\begin{spacing}{1}
		\bibliographystyle{ieeetr}
		\bibliography{mybib}
	\end{spacing}
	

\end{document}